\documentclass[conference]{IEEEtran}
\IEEEoverridecommandlockouts
\usepackage{cite}
\usepackage{amsmath,amssymb,amsfonts}
\usepackage{algorithmic}
\usepackage{graphicx}
\usepackage{textcomp}
\usepackage{xcolor}
\usepackage[hidelinks]{hyperref}
\usepackage{multicol}
\usepackage{lipsum}
\usepackage{array}
\usepackage{tikz}
\usetikzlibrary{arrows.meta}

\def\BibTeX{{\rm B\kern-.05em{\sc i\kern-.025em b}\kern-.08em
    T\kern-.1667em\lower.7ex\hbox{E}\kern-.125emX}}
    
\begin{document}

\title{Non-Invasive Anemia Detection: A Multichannel PPG-Based Hemoglobin Estimation with Explainable Artificial Intelligence\\}

\author{
\IEEEauthorblockN{Garima Sahu, Poorva Verma, Nachiket Tapas}
\IEEEauthorblockA{
\textit{Deptartment of Computer Science and Engineering}\\
\textit{Chhattisgarh Swami Vivekanand Technical University, Bhilai, India}\\
gariimasahu@gmail.com, poorvaverma123@gmail.com, nachikettapas.cse@csvtu.ac.in
}
% \and

% \IEEEauthorblockA{
% \textit{Deptartment of CSE}\\
% \textit{Chhattisgarh Swami Vivekanand}\\
% \textit{Technical University}\\
% Bhilai, India\\
% poorvaverma123@gmail.com}
% \and
% \IEEEauthorblockN{Dr. Nachiket Tapas}
% \IEEEauthorblockA{
% \textit{Deptartment of CSE}\\
% \textit{Chhattisgarh Swami Vivekanand}\\
% \textit{Technical University}\\
% Bhilai, India\\
% nachikettapas.cse@csvtu.ac.in}

}

\maketitle

\begin{abstract}
Anemia is a prevalent hematological disorder that requires frequent hemoglobin monitoring for early diagnosis and effective management. Conventional hemoglobin assessment relies on invasive blood sampling, limiting its suitability for large-scale or continuous screening. This paper presents a non-invasive framework for hemoglobin estimation and anemia screening using multichannel photoplethysmography (PPG) signals and explainable artificial intelligence. Four-wavelength PPG signals (660, 730, 850, and 940~nm) are processed to extract optical and cross-wavelength features, which are aggregated at the subject level to avoid data leakage. A gradient boosting regression model is employed to estimate hemoglobin concentration, followed by post-regression anemia screening using World Health Organization (WHO) thresholds. Model interpretability is achieved using SHapley Additive explanations (SHAP), enabling both global and subject-specific analysis of feature contributions. Experimental evaluation on a publicly available dataset demonstrates a mean absolute error of 8.50 ± 1.27 and a root mean squared error of 8.21~g/L on unseen test subjects, indicating the potential of the proposed approach for interpretable, non-invasive hemoglobin monitoring and preliminary anemia screening.
\end{abstract}

\begin{IEEEkeywords}
Photoplethysmography, Hemoglobin Estimation, Anemia Detection, 
Machine Learning, Explainable Artificial Intelligence
\end{IEEEkeywords}

\section{Introduction}
Anemia is a significant hematological disorder characterized by a decrease in hemoglobin (Hb) concentration, which impairs the oxygen carrying capacity of blood and leads to tissue hypoxia and physiological dysfunction. It remains a major global health concern, affecting more than two billion people worldwide, with a higher prevalence among women, children, and elderly individuals {\cite{1}}. Early detection and effective monitoring of hemoglobin levels are essential to prevent severe health complications such as chronic fatigue, impaired cognitive development, and increased maternal and neonatal morbidity {\cite{2}}.

The conventional diagnostic procedure for anemia involves invasive venous blood collection followed by laboratory-based spectrophotometric analysis for hemoglobin quantification. Although this method provides clinically reliable results, it presents several drawbacks, including patient discomfort, infection risk, dependence on laboratory infrastructure, and limited feasibility for large-scale or frequent screening{\cite{3}}. These limitations highlight the need for an accurate, affordable, and non-invasive technique for hemoglobin estimation that can be utilized in both clinical and community health settings.

Photoplethysmography (PPG) has become a valuable optical sensing technique for non-invasive physiological monitoring. PPG detects pulsatile changes in blood volume by illuminating vascular tissue with light and measuring variations in the reflected or transmitted signal{\cite{4}}. The operation of PPG is based on the Beer–Lambert absorption law, which links the intensity of light passing through biological tissue to the concentration of absorbing chromophores like oxyhemoglobin and deoxyhemoglobin (Hb) {\cite{5}}. Because hemoglobin has wavelength-dependent absorption properties across the visible and near-infrared (NIR) spectrum, multi-wavelength PPG systems can gather detailed optical information about blood composition {\cite{6}}.

Multichannel PPG sensing, using wavelengths such as 660 nm, 730 nm, 850 nm, and 940 nm, enables better discrimination of hemoglobin dynamics and enhances estimation precision under varying physiological and environmental conditions {\cite{7}}. A multi-wavelength sensing architecture can also mitigate challenges arising from tissue scattering, skin pigmentation, and motion artifacts, improving overall signal reliability.

The increasing integration of machine learning (ML) with optical biosignals has expanded the potential of PPG-based systems for hemoglobin estimation. ML algorithms are capable of learning complex, nonlinear relationships between the extracted signal features and physiological parameters, enabling data-driven estimation of hemoglobin concentration {\cite{8, 9}}. While such computational models have shown considerable promise, their limited interpretability poses a challenge for clinical validation and trust in decision support systems {\cite{10}}.

To enhance transparency in algorithmic decision-making, Explainable Artificial Intelligence (XAI) techniques have been introduced in biomedical data analysis. XAI frameworks, such as SHapley Additive exPlanations (SHAP), allow the interpretation of feature importance and facilitate clinical understanding of model behavior .

This work presents a non-invasive hemoglobin estimation framework that combines multichannel PPG sensing, machine learning regression, and explainable AI-based interpretability. The proposed system utilizes four optical wavelengths (660, 730, 850, and 940 nm) to capture wavelength-dependent optical responses of hemoglobin and employs light gradient-boosted decision trees for Hb estimation. The integration of SHAP analysis provides physiological interpretability, ensuring transparency in model predictions. The overall objective is to develop a reliable, low-cost, and explainable framework for non-invasive anemia detection and hemoglobin monitoring applicable to clinical and remote healthcare environments.

\renewcommand{\arraystretch}{1.3}

\begin{table}[htbp]
\caption{Hemoglobin Thresholds for Anemia Classification}
\label{tab:anemia_thresholds}
\centering
\resizebox{\columnwidth}{!}{%
\begin{tabular}{|l|c|c|c|}
\hline
\textbf{Population} & \textbf{Mild Anemia (g/dL)} & \textbf{Moderate Anemia (g/dL)} & \textbf{Severe Anemia (g/dL)} \\
\hline
Children (6--59 months) & 10.0--10.9 & 7.0--9.9 & $<7.0$ \\
Children (5--11 years) & 11.0--11.4 & 8.0--10.9 & $<8.0$ \\
Children (12--14 years) & 11.0--11.9 & 8.0--10.9 & $<8.0$ \\
Non-pregnant Women (15+ yrs) & 11.0--11.9 & 8.0--10.9 & $<8.0$ \\
Pregnant Women & 10.0--10.9 & 7.0--9.9 & $<7.0$ \\
\hline
\end{tabular}
}
\end{table}

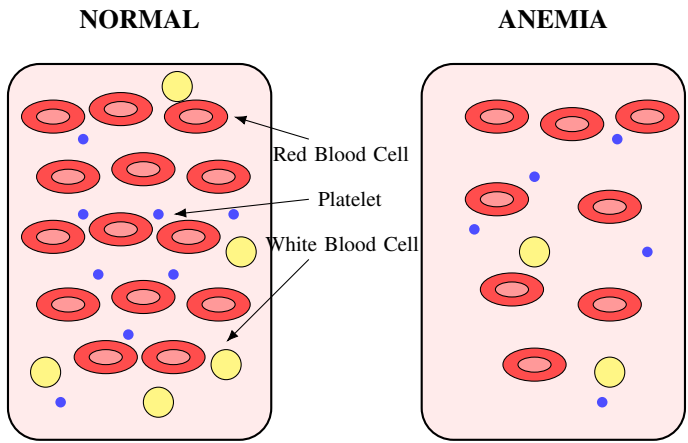
\begin{figure}[htbp]
\centering
\begin{tikzpicture}[scale=1]

% ---------- Titles ----------
\node[font=\bfseries] at (-4.25,5.6) {NORMAL};
\node[font=\bfseries] at (1.25,5.6) {ANEMIA};

% ---------- Containers ----------
\draw[rounded corners=12pt, thick, fill=red!8] (-6,0) rectangle (-2.5,5);
\draw[rounded corners=12pt, thick, fill=red!8] (-0.5,0) rectangle (3,5);

% ---------- RBC macro (biconcave) ----------
\newcommand{\rbc}[2]{
  \draw[fill=red!70] (#1,#2) ellipse (0.42 and 0.22);
  \draw[fill=red!40] (#1,#2) ellipse (0.22 and 0.10);
}

% ---------- RBCs (Normal: many, spaced) ----------
\foreach \x/\y in {
-5.4/4.3,-4.5/4.4,-3.5/4.3,
-5.2/3.5,-4.2/3.6,-3.2/3.5,
-5.4/2.7,-4.5/2.8,-3.6/2.7,
-5.2/1.8,-4.2/1.9,-3.2/1.8,
-4.7/1.1,-3.8/1.1,
}{
\rbc{\x}{\y}
}

% ---------- RBCs (Anemia: fewer) ----------
\foreach \x/\y in {
0.5/4.3,1.5/4.2,2.5/4.3,
0.5/3.2,2/3.1,
2/1.8,0.7/2
}{
\rbc{\x}{\y}
}

% ---------- White Blood Cells (more, non-overlapping) ----------
\draw[fill=yellow!60] (-3.75,4.7) circle (0.20);
\draw[fill=yellow!60] (-3.1,1) circle (0.20);
\draw[fill=yellow!60] (-2.9,2.5) circle (0.20);
\draw[fill=yellow!60] (-4,0.5) circle (0.20);
\draw[fill=yellow!60] (-5.5,0.9) circle (0.20);

\draw[fill=yellow!60] (1,2.5) circle (0.20);
\draw[fill=yellow!60] (2,0.9) circle (0.20);

% ---------- Platelets (more, tiny dots, spaced) ----------
\foreach \x/\y in {
-5.0/3.0,-4.0/3.0,-3.0/3.0,
-4.8/2.2,-3.8/2.2,
-4.4/1.4,-5.0/4.0,-5.3/0.5
}{
  \fill[blue!70] (\x,\y) circle (0.07);
}

\foreach \x/\y in {
2.1/4.0,1/3.5,0.2/2.8,
2.5/2.5,1.9/0.5
}{
  \fill[blue!70] (\x,\y) circle (0.07);
}

% ---------- Labels ----------
\node[anchor=west,font=\footnotesize] at (-2.6,3.8) {Red Blood Cell};
\draw[-{Latex}] (-2,4.0) -- (-3,4.3);

\node[anchor=west,font=\footnotesize] at (-2,3.2) {Platelet};
\draw[-{Latex}] (-2.0,3.2) -- (-3.8,3.0);

\node[anchor=west,font=\footnotesize] at (-2.7,2.6) {White Blood Cell};
\draw[-{Latex}] (-2.0,2.4) -- (-3.1,1.3);

\end{tikzpicture}
\caption{Comparison of normal and anemic blood samples. Anemia is characterized by a reduced concentration of red blood cells.}
\label{fig:normal_vs_anemia}
\end{figure}

\section{Related Work}
Non-invasive hemoglobin (HB) estimation has been actively investigated as an alternative to invasive laboratory-based blood analysis. Optical sensing techniques, particularly photoplethysmography (PPG), have been widely adopted due to their ability to capture pulsatile blood volume changes in peripheral microvascular tissue. The feasibility of PPG based hemoglobin estimation relies on the wavelength-dependent optical absorption properties of hemoglobin chromophores across visible and near-infrared (NIR) spectral regions{\cite{3,5}}.

Several studies have focused on the design of multi-wavelength and multi-site PPG acquisition systems to improve measurement robustness and signal quality. Multi-channel PPG architectures incorporating red and NIR wavelengths have demonstrated improved resistance to motion artifacts, tissue scattering, and skin pigmentation variability {\cite{4,7}}. Spectral-analysis-based approaches have further shown that extending measurements into the visible wavelength range enhances sensitivity to hemoglobin concentration by capturing distinct absorption characteristics {\cite{5,6}}. Reflectance-mode multi-wavelength PPG systems have also been explored for wearable and point-of-care applications, enabling continuous and non-invasive monitoring {\cite{7,11}}.

Beyond sensor design, signal processing and machine learning (ML) techniques have played a central role in interpreting PPG signals for hemoglobin estimation. Supervised ML models have been employed to learn nonlinear relationships between PPG-derived features and laboratory hemoglobin values {\cite{8,9}}. Ensemble-based methods, including Extreme Learning Machines and gradient-boosted models, have demonstrated improved generalization and stability when trained on multi-wavelength PPG data{\cite{9}}. In addition, graph-based and clustering approaches have been proposed to capture temporal and structural characteristics of PPG waveforms, enabling alternative representations for hemoglobin estimation {\cite{6}}.

Deep learning approaches have also been applied to non-invasive Hemoglobin estimation. Convolutional neural networks (CNN) have been used with multi-wavelength reflectance mode PPG signals to enable automatic feature extraction from raw data {\cite{7}}. While such approaches reduce dependence on handcrafted features, they often require large datasets and offer limited interpretability, which restricts their clinical adoption.

Several studies have addressed anemia detection as a classification problem rather than continuous hemoglobin estimation. These frameworks integrate PPG signals with clinical, demographic, or hematological parameters to classify anemic and non-anemic conditions {\cite{2,10}}. IoT-enabled systems have further combined non-invasive sensors with wireless communication to support remote anemia monitoring and large-scale screening {\cite{12}}.

In addition to PPG-based methods, imaging-based techniques have been explored for non-invasive anemia assessment. Retinal image analysis using deep learning models has been proposed to estimate hemoglobin concentration by learning vascular and colorimetric features from fundus images{\cite{13}}. Although promising, such approaches require specialised imaging equipment and are less suitable for continuous monitoring scenarios.

Despite the progress demonstrated in prior studies, challenges remain related to signal variability, model interpretability, and unified system design. Most existing approaches address either sensing or modeling in isolation, with limited integration of explainable frameworks. These limitations motivate the present work, which integrates multichannel PPG sensing, machine learning regression, and explainable artificial intelligence to develop a transparent and reliable framework for non-invasive hemoglobin estimation and anemia detection.

\section{Methodology}
This section describes the proposed framework for non-invasive hemoglobin estimation and anemia screening using multichannel photoplethysmography (PPG) signals. The methodology focuses on signal preprocessing, feature extraction, subject-level regression modeling, and explainable artificial intelligence (XAI)–based interpretation. Anemia screening is implemented as a downstream clinical decision process using World Health Organization (WHO) hemoglobin thresholds applied to the predicted hemoglobin values.

\begin{figure}[htbp]
\centering
\includegraphics[width=\columnwidth]{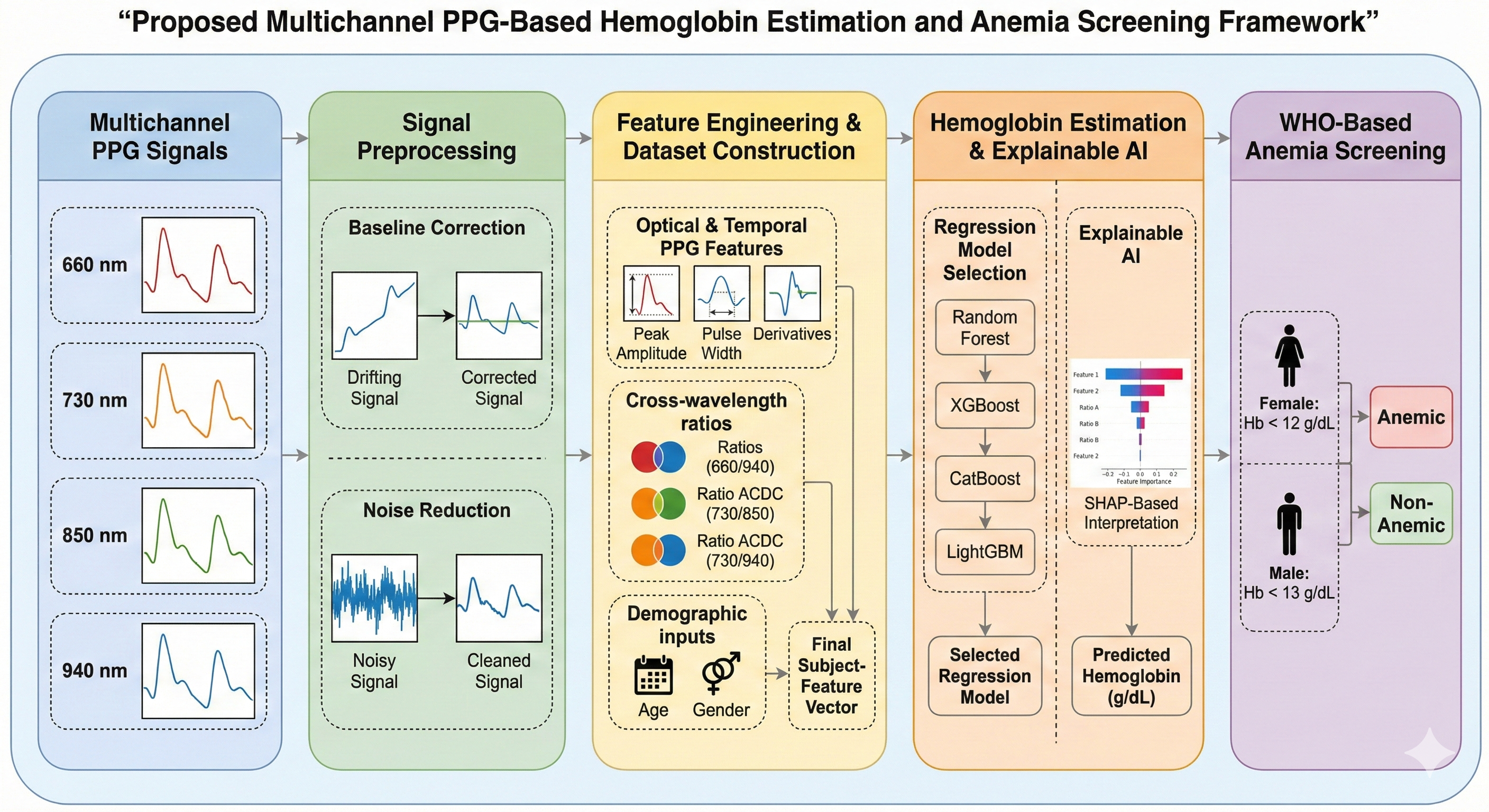}
\caption{Overview of the proposed multichannel PPG-based hemoglobin estimation and WHO-based anemia screening framework.}
\label{fig:pipeline}
\end{figure}

\subsection{Dataset Description}

The proposed framework was evaluated using a publicly available multichannel photoplethysmography (PPG) dataset introduced by Liang Yongbo~\cite{14}. The dataset provides four-wavelength PPG signals acquired at 660~nm, 730~nm, 850~nm, and 940~nm, selected to capture wavelength-dependent optical absorption characteristics of hemoglobin in the visible and near-infrared spectrum. 

The dataset consists of subject-wise recordings accompanied by corresponding reference hemoglobin (Hb) values obtained using standard clinical measurement procedures. In addition to the PPG signals, subject-level metadata such as demographic information are provided, enabling physiologically informed modeling. All subjects in the dataset correspond to 152 adult participants.

In this study, the dataset was organized at the subject level to ensure clinical relevance and to prevent information leakage during model development and evaluation. The publicly available nature of the dataset facilitates reproducibility and supports transparent benchmarking of non-invasive hemoglobin estimation approaches.

\subsection{Signal Preprocessing and Quality Assessment}

Raw multichannel photoplethysmography (PPG) signals were initially inspected in their native form to assess baseline characteristics and noise behavior across all four wavelengths (660, 730, 850, and 940~nm).

Each PPG recording was segmented into fixed-length windows of 500 samples , enabling localized analysis of signal quality and feature stability. Prior to preprocessing, signal quality was quantitatively evaluated using both the signal-to-noise ratio (SNR) and a power spectral density–based signal quality index (SQI). SNR and SQI were computed for every segment and wavelength using Welch’s method, where the physiological PPG band (0.5–5~Hz) was treated as the signal band and all remaining frequencies as noise.

To suppress baseline drift and high-frequency artifacts while preserving the pulsatile component of the PPG waveform, a third-order Butterworth bandpass filter with cut-off frequencies of 0.5–5~Hz was applied independently to each segment and wavelength channel. Following filtering, SNR and SQI were recomputed for all segments to quantitatively verify preprocessing effectiveness. A consistent improvement in both SNR and SQI was observed across wavelengths, confirming that preprocessing enhanced signal fidelity without distorting physiological morphology.
\begin{figure}[htbp]
\centering
\includegraphics[width=\columnwidth]{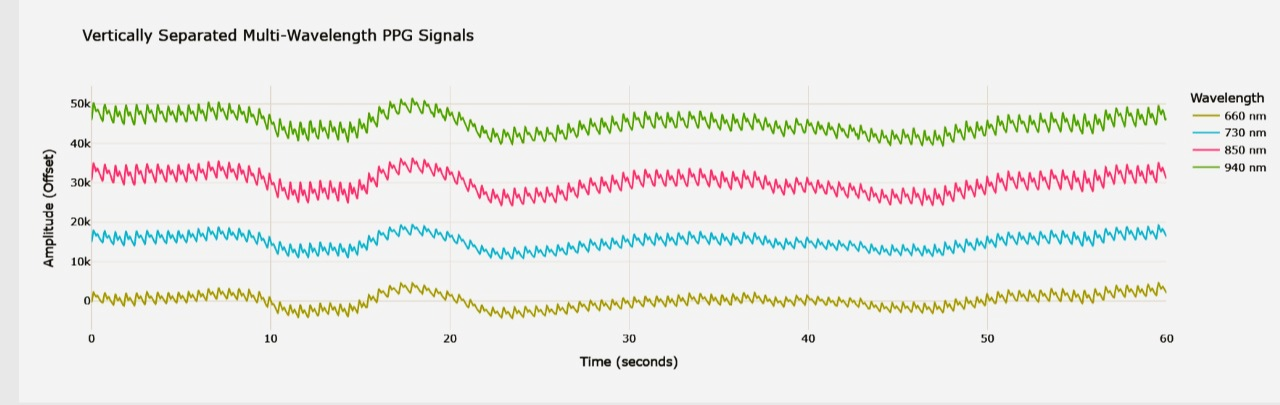}
\caption{Raw PPG Signal}
\label{fig:pipeline}
\end{figure}

\begin{figure}[htbp]
\centering
\includegraphics[width=\columnwidth]{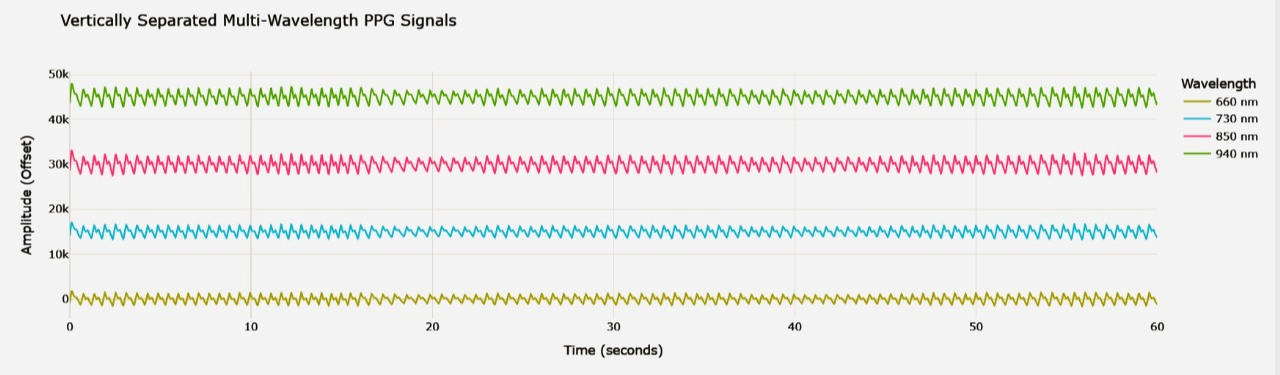}
\caption{Preprocessed PPG Signal}
\label{fig:pipeline}
\end{figure}

\begin{figure}[htbp]
\centering
\includegraphics[width=\columnwidth]{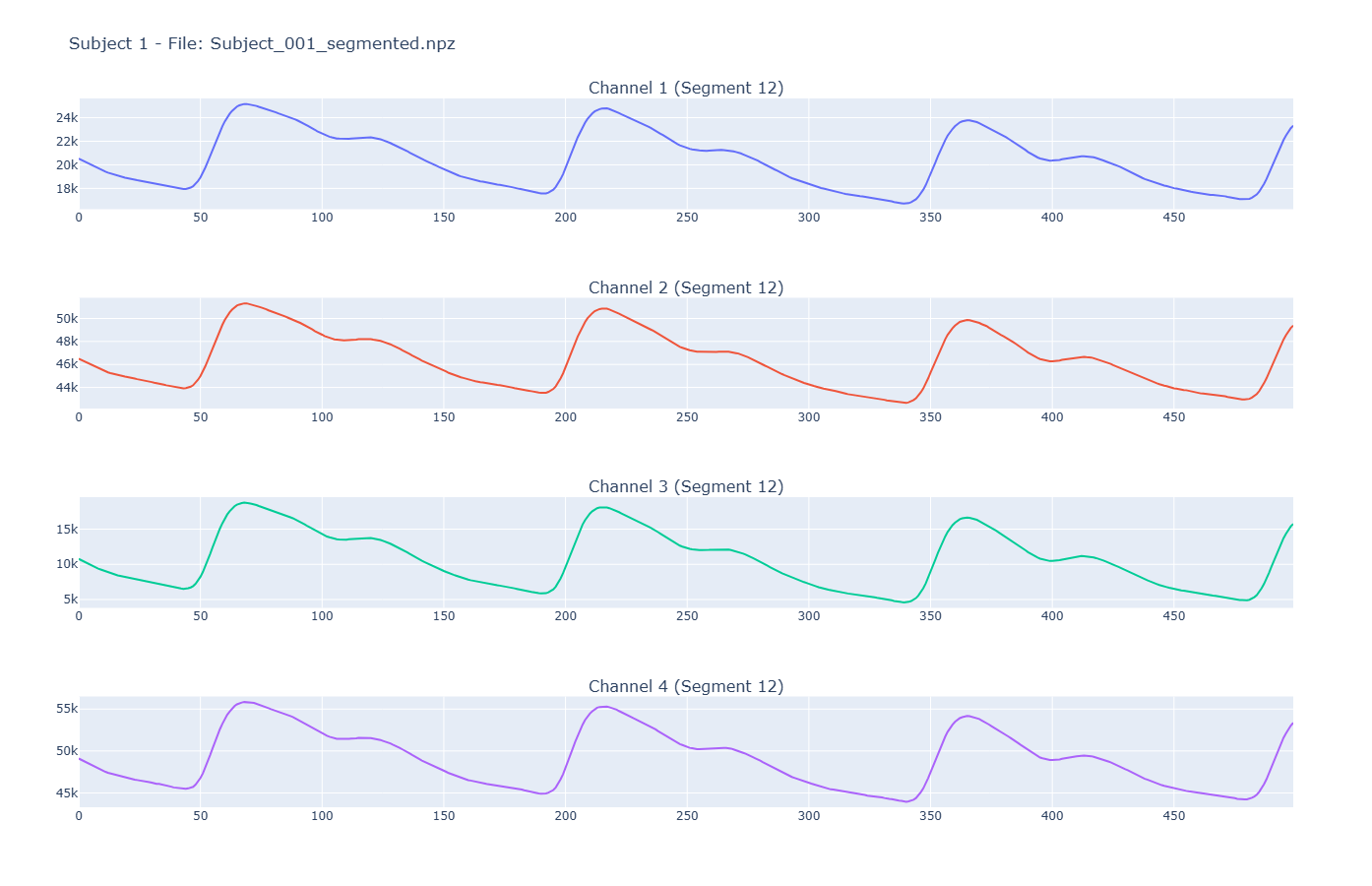}
\caption{Segmented PPG Signal}
\label{fig:pipeline}
\end{figure}
Segment-wise filtering ensured temporal consistency and prevented information leakage across segments. Only filtered signals were retained for subsequent feature extraction.

\subsection{Feature Extraction and Dataset Construction}
From the preprocessed PPG segments, a comprehensive set of physiologically motivated features was extracted independently for each wavelength channel. These features were designed to capture pulsatile amplitude characteristics, baseline absorption behavior, and spectral properties associated with hemoglobin concentration.   

For each segment and wavelength, time-domain statistical features (mean, standard deviation, peak-to-peak amplitude, variance, root mean square, and signal energy) were computed. Pulsatile features were further quantified using alternating current (AC) and direct current (DC) components derived from detected systolic peaks and diastolic troughs, along with their normalized ratio (AC/DC). Logarithmic attenuation features were additionally calculated to represent relative absorption behavior.

To incorporate spectral information, frequency-domain features including dominant frequency, spectral entropy, and normalized band power within the physiological heart-rate range were extracted using Welch’s power spectral density estimation. Cross-wavelength ratios were subsequently computed for selected feature groups (mean, AC/DC, and attenuation) to encode relative optical absorption differences between wavelength pairs.

Each segment-level feature vector was associated with the corresponding subject hemoglobin value and demographic information (age and gender), forming an initial machine-learning-ready dataset. Feature tables were subjected to cleaning procedures, including removal of NaN-heavy and constant columns, and exclusion of unrelated clinical measurements. This process resulted in a stable and interpretable multichannel PPG feature set.The extracted feature categories and their mathematical definitions are summarized in Table~\ref{tab:features}.

\begin{table}[htbp]
\caption{Summary of Extracted Multichannel PPG Features}
\label{tab:features}
\centering
\resizebox{\columnwidth}{!}{%
\begin{tabular}{|c|l|l|l|}
\hline
\textbf{Category} & \textbf{Feature} & \textbf{Definition} & \textbf{Description} \\
\hline
Time-domain &
Mean &
$\mu = \frac{1}{N}\sum_{i=1}^{N} x_i$ &
Average signal amplitude \\

&
Standard Deviation &
$\sigma = \sqrt{\frac{1}{N}\sum(x_i-\mu)^2}$ &
Signal variability \\

&
RMS &
$\sqrt{\frac{1}{N}\sum x_i^2}$ &
Signal energy estimate \\

&
Peak-to-Peak &
$\max(x)-\min(x)$ &
Pulse amplitude variation \\
\hline

Optical &
AC &
$A_{\text{sys}} - A_{\text{dia}}$ &
Pulsatile component \\

&
DC &
$\mu(x)$ &
Baseline absorption \\

&
AC/DC &
$\frac{AC}{DC}$ &
Normalized pulsatile index \\
\hline

Spectral &
Dominant Frequency &
$\arg\max(P(f))$ &
Heart-rate related component \\

&
Band Power &
$\sum_{f_1}^{f_2} P(f)$ &
Physiological power content \\

&
Spectral Entropy &
$-\sum p(f)\log p(f)$ &
Spectral complexity \\
\hline

Cross-wavelength &
Feature Ratio &
$\frac{F_{\lambda_i}}{F_{\lambda_j}}$ &
Relative optical absorption \\

&
Attenuation Ratio &
$\log\left(\frac{I_{\lambda_i}}{I_{\lambda_j}}\right)$ &
Hemoglobin-sensitive contrast \\
\hline

Aggregation &
Mean / Median &
$\text{stat}(F_s)$ &
Subject-level aggregation \\
\hline
\end{tabular}
}
\end{table}

\subsection{Hemoglobin Estimation and Model Training}

Hemoglobin estimation was formulated as a supervised regression problem, where the objective was to predict continuous hemoglobin concentration from multichannel PPG-derived features. Initial experiments were conducted using segment-level feature representations to enable the models to learn local relationships between PPG waveform characteristics and hemoglobin values. However, because multiple segments originating from the same subject shared an identical reference hemoglobin value, this formulation introduced redundancy and increased susceptibility to overfitting.

To address this limitation and align the learning process with clinical interpretation, subject-level feature aggregation was subsequently performed. For each subject, segment-level features were aggregated using robust statistical operators, including mean and median, resulting in a single representative feature vector per subject. This aggregation reduced label duplication, improved generalization, and ensured that model inputs corresponded directly to subject-level hemoglobin measurements.

Multiple regression models were evaluated, including Random Forest, Cat Boost, XGBoost, and Light GBM. Light GBM provided the most stable performance, achieving a favorable balance between predictive accuracy and generalization. Consequently, it was selected as the final hemoglobin estimation model.

Model training and evaluation were performed using a subject-wise data partitioning strategy, ensuring that no subject appeared in both training and test sets. An 80:20 subject-level split was employed, with the held-out test set reserved exclusively for final performance evaluation.

\subsection{Explainable AI and WHO-Based Anemia Screening}

To enhance transparency and clinical interpretability of the hemoglobin estimation model, explainable artificial intelligence (XAI) analysis was conducted using SHapley Additive exPlanations (SHAP). SHAP analysis was applied to the final LightGBM regression model to quantify the contribution of individual input features to predicted hemoglobin values. Both model-intrinsic gain-based feature importance and SHAP-based explanations were examined to capture complementary perspectives on feature relevance.

Global interpretability was assessed using SHAP summary and bar plots, highlighting the relative influence of multichannel optical features, cross-wavelength ratios, and demographic variables on hemoglobin prediction. In addition, local interpretability was explored through SHAP waterfall plots for individual test subjects, enabling subject-specific analysis of feature contributions. SHAP dependence plots were further used to investigate nonlinear relationships between selected optical features and model outputs.

Anemia screening was subsequently performed as a post-regression decision step using World Health Organization (WHO) hemoglobin thresholds. Predicted hemoglobin values obtained from the regression model were mapped to anemia status using sex-specific criteria (Hb~$<$~130 g/L for males and Hb~$<$~120 g/L for females). No classification model was trained, and anemia labels were not used during model learning, thereby avoiding circular inference. Since ground-truth anemia annotations were unavailable, classification performance metrics were not reported; instead, threshold-based screening outcomes and sensitivity to threshold offsets were analyzed to assess robustness of the decision boundaries.

\section{Results and Discussion}

\subsection{Hemoglobin Estimation Performance}

The performance of the proposed multichannel PPG-based hemoglobin estimation framework was evaluated at the subject level using mean absolute error (MAE), root mean squared error (RMSE), and coefficient of determination ($R^2$). 
The LightGBM regression model demonstrated strong learning capability, achieving a training MAE of approximately 2--3~g/L and an RMSE of 3.82~g/L. On unseen test subjects, the model attained an MAE  8--9~g/L, an RMSE of 8.21~g/L. The gap between training and test performance reflects the inherent challenges of non-invasive optical hemoglobin estimation, including inter-subject physiological variability and signal attenuation effects.
\begin{figure}[htbp]
\centering
\includegraphics[width=\columnwidth]{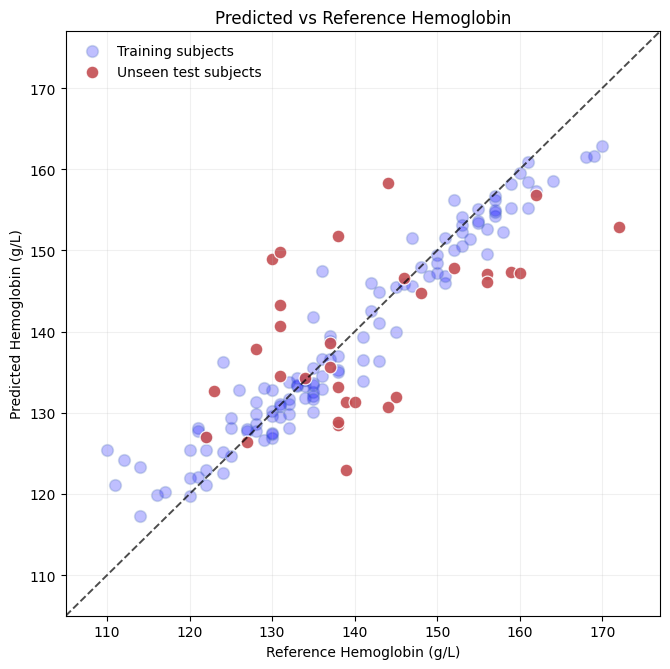}
\caption{Predicted versus Actual hemoglobin values }
\label{fig:hb_scatter}
\end{figure}

\begin{figure}[htbp]
\centering
\includegraphics[width=\columnwidth]{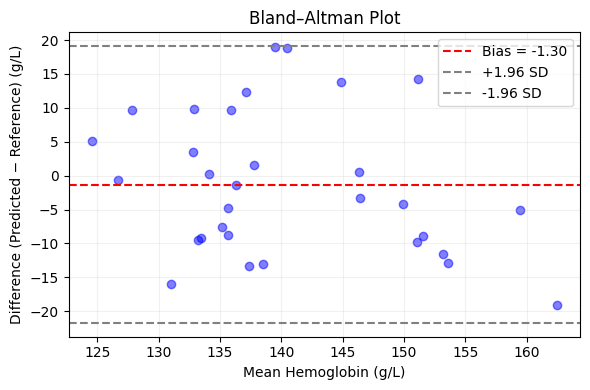}
\caption{Bland--Altman plot comparing predicted and reference hemoglobin values for test subjects.}
\label{fig:hb_ba}
\end{figure}

Figure~\ref{fig:hb_scatter} shows the predicted versus reference hemoglobin values for both training and test subjects. Training samples are included to illustrate the learned regression trend, while test samples highlight the generalization behavior of the model. Despite increased dispersion at higher hemoglobin values, the overall trend follows the line of identity.Agreement between predicted and reference hemoglobin measurements on the test set was further examined using Bland--Altman analysis (Fig.~\ref{fig:hb_ba}). The mean bias was observed to be small, with most samples lying within the $\pm1.96$ standard deviation limits, indicating the absence of strong systematic bias.

\subsection{Model Interpretability Analysis}

To improve transparency and clinical interpretability of the hemoglobin estimation model, explainable artificial intelligence (XAI) analysis was performed using SHapley Additive exPlanations (SHAP). SHAP was applied to the final LightGBM regression model to quantify the contribution of individual input features toward predicted hemoglobin values, enabling both global and subject-level interpretation.

\begin{figure}[htbp]
\centering
\includegraphics[width=\columnwidth]{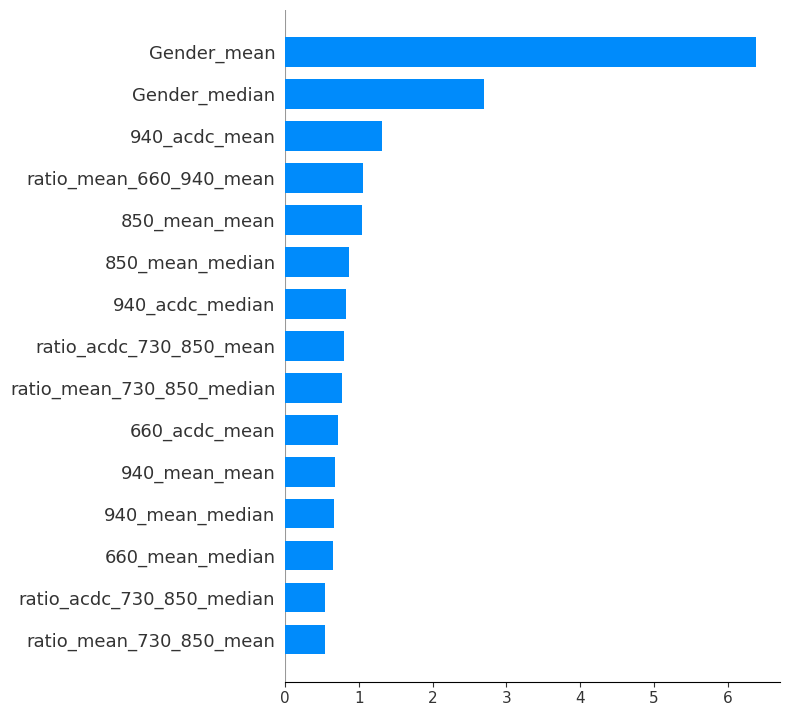}
\caption{Global SHAP feature importance based on mean absolute SHAP values, highlighting the relative contribution of features and demographic variables to hemoglobin prediction.}
\label{fig:shap_global}
\end{figure}

Figure~\ref{fig:shap_global} presents the global SHAP summary analysis, illustrating the relative importance of multichannel optical features, cross-wavelength ratios, and demographic variables. Features derived from near-infrared wavelengths and AC/DC-related ratios consistently exhibited strong contributions, reflecting the wavelength-dependent absorption characteristics of hemoglobin. Demographic information, particularly gender, also influenced predictions, which is physiologically consistent with known differences in hemoglobin reference ranges.

Local interpretability was further examined using SHAP waterfall plots for individual test subjects. These visualizations highlight how multiple features interact nonlinearly to increase or decrease the predicted hemoglobin value for a given subject. 

The XAI analysis confirms that the regression model learns interpretable and physiologically plausible relationships rather than acting as a black-box predictor.

\subsection{WHO-Based Anemia Screening Outcomes}

The screening analysis focused on the distribution of predicted anemia status and the robustness of threshold-based decisions. The majority of test subjects were classified as non-anemic under standard WHO thresholds, consistent with the hemoglobin distribution observed in the dataset.

Since ground-truth anemia labels were not available, conventional classification performance metrics such as accuracy, sensitivity, or specificity were not computed. Instead, the screening analysis focused on the plausibility and robustness of threshold-based decisions.To assess sensitivity to decision boundaries, threshold offset analysis was conducted by systematically shifting the WHO thresholds. The predicted number of anemic cases remained stable under moderate threshold perturbations, indicating robustness of the screening outcomes with respect to small estimation errors. These findings suggest that regression-based hemoglobin estimation followed by rule-based screening may serve as a practical approach for preliminary anemia risk assessment, particularly in resource-constrained or non-invasive monitoring settings.

\section{Conclusion}

This paper presented a non-invasive framework for hemoglobin estimation and anemia screening using multichannel photoplethysmography (PPG) signals and explainable machine learning. By leveraging four-wavelength PPG measurements, subject-level feature aggregation, and gradient boosting regression, the proposed approach enables hemoglobin estimation without invasive blood sampling while maintaining model transparency through SHAP-based interpretability. Post-regression anemia screening using World Health Organization (WHO) thresholds aligns the framework with real-world clinical decision-making . Experimental results on a public dataset demonstrate that multichannel PPG contains physiologically meaningful information for hemoglobin estimation, though residual variability remains due to inter-subject differences and optical signal limitations. Future work will focus on expanding the dataset to include broader clinical populations such as pediatric subjects and pregnant women, incorporating clinically labeled anemia severity levels (mild, moderate, and severe), and exploring advanced feature representations and learning strategies to further improve robustness and generalization.

%IEEEabrv,
\bibliographystyle{IEEEtran}
\bibliography{bibliography}

@misc{1,
	author = {},
	title = {{A}naemia in women and children --- who.int},
	howpublished = {\url{https://www.who.int/data/gho/data/themes/topics/anaemia_in_women_and_children}},
	year = {},
	note = {[Accessed 20-01-2026]},
}

@INPROCEEDINGS{2,
  author={Bahadure, Nilesh Bhaskarrao and Khomane, Ramdas and Nittala, Aditya and Singh, Abhinav and Chadalwada, Dev and Bonde, Om and Nagar, Somesh and Raju, Nagrajan},
  booktitle={2023 International Conference on Artificial Intelligence for Innovations in Healthcare Industries (ICAIIHI)}, 
  title={Anemia Detection and Classification Using Data Analysis of Blood Samples}, 
  year={2023},
  doi={10.1109/ICAIIHI57871.2023.10489465}}

@article{3,
author = {Kumar, R and Guruprasad, S and Kansara, Krity and Rao, K and Mohan, Murali and Reddy, Manjunath and Prabhu, Uday and Prakash, P and Chakraborty, Sushovan and Das, Sreetama and Madhusoodanan, K.N.},
year = {2021},
title = {A Novel Noninvasive Hemoglobin Sensing Device for Anemia Screening},
journal = {IEEE Sensors Journal},
doi = {10.1109/JSEN.2021.3070971}
}

@ARTICLE{4,
  author={Karolcik, Stefan and Ming, Damien K. and Yacoub, Sophie and Holmes, Alison H. and Georgiou, Pantelis},
  journal={IEEE Transactions on Biomedical Circuits and Systems}, 
  title={A Multi-Site, Multi-Wavelength PPG Platform for Continuous Non-Invasive Health Monitoring in Hospital Settings}, 
  year={2023},
  doi={10.1109/TBCAS.2023.3254453}}

@article{5,
author = {Wang, Yunyi and Li, Gang and Kong, Li and Lin, Ling},
year = {2023},
month = {09},
title = {High-precision non-invasive RBC and HGB detection system based on spectral analysis},
journal = {Analytical and Bioanalytical Chemistry},
doi = {10.1007/s00216-023-04950-x}
}

@Article{6,
AUTHOR = {Liu, Lei and Wang, Ziyi and Zhang, Xiaohan and Zhuang, Yan and Liang, Yongbo},
TITLE = {A Novel Model for Noninvasive Haemoglobin Detection Based on Visibility Network and Clustering Network for Multi-Wavelength PPG Signals},
JOURNAL = {Algorithms},
YEAR = {2025},
URL = {https://www.mdpi.com/1999-4893/18/2/75},
ISSN = {1999-4893},
DOI = {10.3390/a18020075}
}

@INPROCEEDINGS{7,
  author={Lychagov, Vladislav and Semenov, Vladimir and Volkova, Elena and Chernakov, Dmitrii and Ahn, Joongwoo and Kim, Justin Younghyun},
  booktitle={2023 45th Annual International Conference of the IEEE Engineering in Medicine and Biology Society (EMBC)}, 
  title={Non-invasive hemoglobin concentration measurements with multi-wavelength reflectance mode PPG sensor and CNN data processing*}, 
  year={2023},
  doi={10.1109/EMBC40787.2023.10341173}}

@INPROCEEDINGS{8,
  author={M, Jayasanthi and V, Vismitha and Joshi A P, Eunice},
  booktitle={2024 International Conference on Smart Systems for Electrical, Electronics, Communication and Computer Engineering (ICSSEECC)}, 
  title={A Non-Invasive Approach for Hemoglobin Estimation}, 
  year={2024},
  doi={10.1109/ICSSEECC61126.2024.10649469}}

@article{9,
author = {Fulai, Peng and Zhang, Ningling and Chen, Cai and Wu, Fengxia and Wang, Weidong},
year = {2024},
title = {Ensemble Extreme Learning Machine Method for Hemoglobin Estimation Based on PhotoPlethysmoGraphic Signals},
journal = {Sensors},
doi = {10.3390/s24061736}
}

@article{10,
author = {Saleh, Neven and Salaheldin, Ahmed and Ismail, Yasser and Afify, Heba},
year = {2025},
title = {Classification of anemic condition based on photoplethysmography signals and clinical dataset},
journal = {Biomedical Engineering / Biomedizinische Technik},
doi = {10.1515/bmt-2024-0433}
}

@INPROCEEDINGS{11,
  author={Abhishek, Kumar and Saxena, Amodh Kant and Sonkar, Ramesh Kumar},
  booktitle={2015 IEEE International Conference on Signal Processing, Informatics, Communication and Energy Systems (SPICES)}, 
  title={Non-invasive measurement of heart rate and hemoglobin concentration level through fingertip}, 
  year={2015},
  doi={10.1109/SPICES.2015.7091549}}

@article{12,
author = {Ravi, Abiramee and Priyadarshini, S. and Dharmarajan, Krithina and Jaya, Suriya},
year = {2024},
title = {Diagnosis of anemia using non-invasive anemia detector through parametrical analysis},
journal = {i-manager’s Journal on Wireless Communication Networks},
doi = {10.26634/jwcn.13.1.21118}
}

@article{13,
  title={Noninvasive Anemia Detection and Hemoglobin Estimation from Retinal Images Using Deep Learning: A Scalable Solution for Resource-Limited Settings},
  author={Khan, Rehana and Maseedupally, Vinod and Thakoor, Kaveri A and Raman, Rajiv and Roy, Maitreyee},
  journal={Translational Vision Science \& Technology},
  year={2025},
  publisher={The Association for Research in Vision and Ophthalmology}
}

@article{14,
author = "Liang yongbo",
title = "{Non-invasive Hemoglobin Detection based on Four-wavelength PPG Signal}",
year = "2023",
month = "3",
url = "https://figshare.com/articles/dataset/Hemoglobin_detection_based_on_four-wavelength_PPG_signal_zip/22256143",
doi = "10.6084/m9.figshare.22256143.v5"
}
\end{document}